# Towards Secure Logging: Characterizing and Benchmarking Logging Code Security Issues with LLMs

HE YANG YUAN*, York University, Canada
XIN WANG*, The Hong Kong University of Science and Technology (Guangzhou), China
KUNDI YAO, University of Waterloo, Canada
AN RAN CHEN, University of Alberta, Canada
ZISHUO DING†, The Hong Kong University of Science and Technology (Guangzhou), China
ZHENHAO LI†, York University, Canada

Logging code plays an important role in software systems by recording key events and behaviors, which are essential for debugging and monitoring. However, insecure logging practices can inadvertently expose sensitive information or enable attacks such as log injection, posing serious threats to system security and privacy. Prior research has examined general defects in logging code, but systematic analysis of logging code security issues remains limited, particularly in leveraging LLMs for detection and repair. In this paper, we derive a comprehensive taxonomy of logging code security issues, encompassing four common issue categories and 10 corresponding patterns. We further construct a benchmark dataset with 101 real-world logging security issue reports that have been manually reviewed and annotated. We then propose an automated framework that incorporates various contextual knowledge to evaluate LLMs' capabilities in detecting and repairing logging security issues. Our experimental results reveal a notable disparity in performance: while LLMs are moderately effective at detecting security issues (e.g., the accuracy ranges from 12.9% to 52.5% on average), they face noticeable challenges in reliably generating correct code repairs. We also find that the issue description alone improves the LLMs' detection accuracy more than the security pattern explanation or a combination of both. Overall, our findings provide actionable insights for practitioners and highlight the potential and limitations of current LLMs for secure logging.

## 1 Introduction

Logging is widely adopted in software development as a way of capturing information about program execution in the form of logs. Developers typically insert logging statements into the

---

* Equal contribution.
† Corresponding authors.

Authors' Contact Information: He Yang Yuan, York University, Toronto, Canada, yuanh@yorku.ca; Xin Wang, The Hong Kong University of Science and Technology (Guangzhou), Guangzhou, China, xwang496@connect.hkust-gz.edu.cn; Kundi Yao, University of Waterloo, Waterloo, Canada, kundi.yao@uwaterloo.ca; An Ran Chen, University of Alberta, Edmonton, Canada, anran6@ualberta.ca; Zishuo Ding, The Hong Kong University of Science and Technology (Guangzhou), Guangzhou, China, zishuoding@hkust-gz.edu.cn; Zhenhao Li, York University, Toronto, Canada, lzhenhao@yorku.ca.

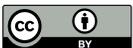

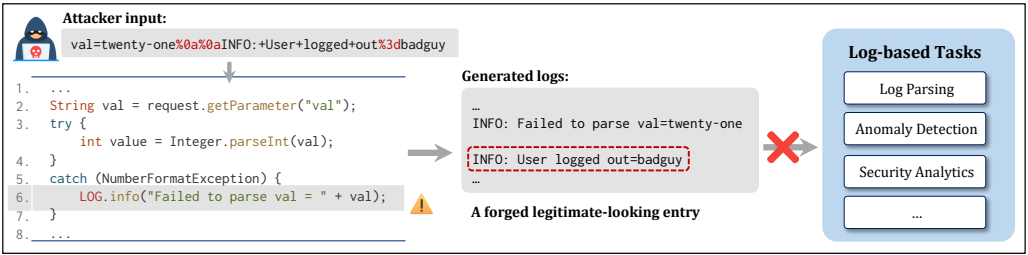

Fig. 1. A log injection vulnerability. Malicious input containing encoded newline characters is logged without sanitization, creating a forged log entry. This may mislead downstream components such as anomaly detectors and security monitors.

codebase, which generate logs at runtime by recording key events, contextual data, and system states [21, 31]. These logs provide developers and system operators with visibility into runtime behavior and serve as a foundational source of information for various software engineering tasks. In particular, logs play an essential role in debugging [58], performance analysis [59, 60], system monitoring [10, 24], and security auditing [12]. A typical logging statement includes a severity indicator (such as *"info"* or *"error"*), a descriptive message, and runtime variables that provide contextual details. For example, as shown in Figure 1, a developer may log an error when parsing user input fails, recording both the error message and the input value to aid debugging. The resulting logs allow developers to trace system behavior, diagnose failures, and understand how external inputs interact with the application during execution.

Although logs provide valuable insights into system behavior, improper logging practices can introduce serious security and privacy risks. For example, if logs are generated without appropriate safeguards such as input sanitization, sensitive data redaction, or format validation, they may inadvertently expose confidential information or become vectors for attacks such as log injection (e.g., Log4Shell [2]). As illustrated in Figure 1, unescaped user input embedded within a logging statement can be crafted to inject forged entries that resemble legitimate log messages. As a result, these forged entries may compromise the integrity of the log stream and contaminate downstream log-based tasks, including parsing [27, 28, 38], root cause analysis, and security monitoring.

While prior research has acknowledged the security and privacy risks associated with logs, most existing work has focused primarily on the content of generated logs, especially from a data privacy perspective [6, 7]. However, these postmortem approaches are inherently reactive as they address vulnerabilities only after potentially sensitive data has already been recorded and exposed. More importantly, these approaches often neglect the upstream causes of insecure logging, namely the code and developer decisions that introduce such risks in the first place. Despite the importance of preventing insecure logs at their source, limited research has explored how developers may inadvertently expose sensitive or security-critical information through logging statements.

In addition, several studies have investigated the characteristics of logging statements and developed tools to detect common issues such as missing logs, incorrect severity levels, or inconsistent variable usage [18, 34, 37]. While these efforts enhance the general quality of logging code, they rarely address the security dimension of logging practices. Risks such as log injection or log forging, where attackers influence log content to mislead downstream analysis, have received little attention. As a result, existing detection tools might be insufficient for identifying security-sensitive flaws introduced during logging. These limitations highlight the need for approaches that can reason about both the content and context of logging statements, particularly in security-sensitive scenarios.

To bridge this knowledge gap, we first conduct a systematic empirical study to characterize the landscape of logging security issues. This foundational analysis is essential to foster a comprehensive understanding of logging security issues and how they manifest in real-world code, thereby informing the development of effective detection and mitigation strategies. Through a systematic analysis of existing literature and established vulnerability databases, including the Common Weakness Enumeration (CWE) [3] and the Common Vulnerabilities and Exposures (CVE) [1], we derive a detailed taxonomy that classifies logging security issues into four main categories and 10 specific patterns. This taxonomy serves as the theoretical foundation for understanding the scope and nature of logging security issues. Building on this empirical foundation, we then explore the potential of LLMs to detect and repair the security issues in logging code. LLMs have demonstrated strong capabilities in understanding and reasoning about source code, supporting tasks such as code completion, defect detection, and performance modeling [9, 13, 14, 41, 46, 48, 51, 52, 61]. However, their effectiveness in identifying and repairing logging security issues remains largely unexplored. In this study, we construct a benchmark dataset, **SecLogging**, which contains 101 real-world logging security issues grounded in our taxonomy and systematically evaluate multiple open-source and proprietary LLMs on their ability to detect and repair these real-world issues.

Our evaluation results show that the ability of LLMs to detect logging code security issues varies considerably, with category-level accuracy ranging from 12.9% to 52.5%. The most persistent challenge arises in abstract, logic-intensive cases such as Improper Redaction or Masking (RM), where most LLMs score below 20%. In contrast, concrete categories like Sensitive Information Exposure (SS) are relatively easier to detect, with DeepSeek-V3 reaching as high as 85.7% accuracy under certain settings. Additionally, we find that introducing targeted contextual knowledge, particularly a brief issue description (+D), often helps the models improve, yielding noticeable accuracy gains (e.g., the category accuracy on average increases from 41.6% to 52.5% for DeepSeek-V3). Yet, richer contextual inputs do not guarantee further benefits. In fact, when descriptions and explanations are combined (+E+D), the performance of several models declines, which suggests that excess detail can overwhelm the models rather than guide them effectively. For repair, the challenge becomes more evident. Patch Similarity ranges from 18.8% to 69.3%, but the highest average performance is observed in the Base setting (44.1%) without additional contextual information, whereas the most enriched prompts (+E+D) drop to 36.5% on average. The results indicate that concise prompts support more precise repairs, while overly elaborate context often introduces noise that degrades repair capability.

Overall, our results highlight several key insights in secure logging and improving LLMs. We find that LLMs show uneven effectiveness: while they can detect certain issues with reasonable accuracy, their ability to repair issues is less reliable. Targeted and concise contextual information can improve detection, but excessive detail often harms both detection and repair quality. These findings reveal the importance of precise prompting strategies, cost-efficient contextual enrichment, and continued human oversight.

We summarize the contributions of this paper as follows:

- We derive a comprehensive taxonomy of logging code security issues synthesized from literature and vulnerability databases, which covers four categories and 10 patterns. The taxonomy provides practical insights for practitioners and researchers aiming at secure logging.
- We construct a benchmark dataset, **SecLogging**, consisting of 101 curated logging security issues annotated with our taxonomy. This benchmark enables systematic evaluation of LLMs' capabilities in detecting and repairing real-world logging security issues.

- We conduct a systematic evaluation of LLMs using our benchmark, and uncover actionable insights for practitioners and researchers on secure logging practices as well as directions for improving LLMs in this domain.

**Paper Organization.** The remainder of this paper is organized as follows. Section 2 summarizes the background and related work of this paper. Section 3 details our methodology. Section 4 outlines the experimental setup, covering the chosen LLMs and our evaluation metrics. Section 5 presents the results. We discuss the implications of our findings in Section 6 and acknowledge the study's limitations in Section 7. Finally, Section 8 concludes the paper.

## 2 Background and Related Work

In this section, we first provide background on key concepts essential to our study, including the role of logging and the standardized vulnerability databases we utilize. We then summarize prior research across three related domains: (1) security and privacy concerns specific to software logging, (2) broader challenges surrounding logging code quality, and (3) the emerging application of LLMs for logging and log analysis.

### 2.1 Background

Software logging is a critical mechanism for monitoring, debugging, and auditing applications in production [8, 10, 21, 24, 30, 31, 58]. While indispensable, logging practices can also introduce significant security risks if not implemented carefully. To systematically categorize these risks, the security community relies on standardized references. The CWE is a community-developed list of software and hardware weakness types that serves as a common language for describing security flaws [3, 33, 57]. The CVE database provides a dictionary of publicly known information-security vulnerabilities in specific products, assigning each a unique identifier [1, 20]. Our work leverages both CWE and CVE to ground our taxonomy of logging code security issues in established security standards.

### 2.2 Logging Security and Privacy

The exposure of sensitive information in logs has long been recognized as a critical security and privacy concern. Prior studies have primarily focused on privacy leakage from logs [6]. For example, several works categorized different types of sensitive data disclosure and examined their consequences for users and systems [7, 63, 64, 67]. Other efforts investigated logging-related risks in big data environments [44]. Research has also addressed specific attack vectors, such as log injection, with systems like LogInjector targeting injection flaws in web applications [45]. However, despite these advances, prior work has largely centered on analyzing the information contained in logs, while the security of the logging code itself has not been systematically studied. Our work fills this gap by providing a comprehensive taxonomy and benchmark of logging code security issues.

### 2.3 Logging Code Quality

The quality of logging code directly impacts a system's maintainability and debuggability [50], as common anti-patterns like excessive logging, misused logging levels, and insufficient contextual detail can introduce significant system-level issues. Some studies have investigated general logging-related issues by analyzing issue reports from two projects [23], while others have focused on specific aspects such as "how-to-log" [11] and "where-to-log" practices [35], temporal inconsistencies between logging and surrounding code [19], duplicate logging code [36], and readability issues [34]. These efforts typically rely on mining commit histories [11, 19, 36] or conducting developer surveys [34]. More recently, Zhong et al. [65] proposed LogUpdater, which identifies logging defects based on types mined from commit histories, primarily targeting inconsistency and readability issues. Prior studies have revealed that these logging code quality problems mainly affect the usefulness, reliability, and interpretability of the generated logs themselves. In contrast, the logging code security issues examined in our work can compromise the logs (e.g., leaking sensitive information) and further pose risks to the stability and security of the whole system.

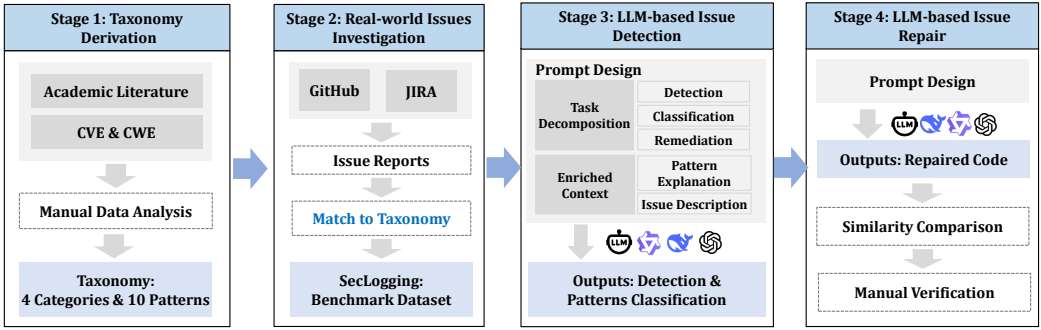

Fig. 2. An overview of our study.

### 2.4 LLMs for Logging and Log Analysis

Researchers have started probing LLM's capabilities for generating logging code. For example, LANCE [42] employs the T5 model to automatically generate complete logging code. Similarly, UniLog leverages in-context learning, enabling LLMs to generate logging code with only a few example prompts, without requiring fine-tuning [55]. Prior studies have employed LLMs to automatically generate complete logging code based on source code context [32, 66]. These approaches demonstrate the potential of LLMs to automate logging practices.

In parallel, LLMs have also been applied to log analysis, particularly log parsing. LogBatcher [54] targets the need for labeled data or model training by clustering logs, matching caches, and applying batch prompt contexts, improving efficiency and scalability. KnowLog enhances [39] pre-trained language models with domain-specific knowledge and contrastive learning to better capture abbreviations and contextual cues in logs, thereby improving exception detection. LibreLog builds on open-source LLMs [40] to perform unsupervised log parsing using similarity scoring, self-reflection, and template recall, which achieved high parsing accuracy and speed. These advances illustrate how LLMs can streamline the tasks of log analysis, especially log parsing.

Building on these lines of research, our work focuses on the security dimension of logging code. We leverage LLMs to detect, classify security issues in logging code, and generate repair solutions. In this way, our study complements prior work on logging automation and log analysis, while filling an important gap in understanding and addressing security issues in logging code.

## 3 Methodology

In this paper, we conduct a comprehensive study on security issues in logging code. Figure 2 presents an overview of our study, which involves four stages. **Stage 1:** We collect and analyze resources related to logging code security issues, including academic literature, CVEs, and CWEs. Based on this analysis, we derive a taxonomy that characterizes common categories and patterns of logging code security issues. **Stage 2:** We collect real-world issue reports from open-source projects and annotate them according to the taxonomy established in Stage 1. **Stage 3:** We investigate the capability of LLMs to automatically detect these security issues by leveraging the constructed taxonomy as guidance. ***Stage 4:*** We further explore the potential of LLMs to repair the detected issues.

### 3.1 Deriving the Taxonomy of Logging Code Security Issues

To construct a comprehensive taxonomy of security issues in logging code, we systematically gather and analyze data from three sources: academic literature, the Common Weakness Enumeration (CWE), and the Common Vulnerabilities and Exposures (CVE) database in Stage 1.

**Academic Literature:** To the best of our knowledge, there are no prior studies that directly focus on security issues in logging code. Therefore, we begin by collecting papers that investigate defects in logging or improving logging practices. Similar to prior studies [15, 25, 29], our literature review follows a structured process to identify relevant research. First, we perform an initial search of major digital libraries (IEEE Xplore, ACM Digital Library, SpringerLink), targeting publications from 2019 to 2025. Our search keywords include "logging", "logging security", "sensitive log", and "log injection". We then screen the titles and abstracts of the retrieved articles to filter out irrelevant publications. Finally, we conduct a full-text review of the remaining papers and follow a snowballing strategy [53], examining their citations to discover additional relevant work. This process yields a list of 111 papers for further analysis.

**CWE and CVE:** To ground our taxonomy in established security standards and real-world incidents, we analyze data from two key vulnerability databases. First, we manually review all entries in the MITRE CWE catalog [3] and identify 13 CWE instances related to logging code based on their descriptions. Second, we search the CVE database [1] using the keywords "logging" and "logs", then manually filter the retrieved results to obtain 19 CVEs that represent concrete cases of logging code security issues.

**Data Analysis.** We first examine the titles and abstracts of all collected papers and filter out those unrelated to security issues in logging code. Examples of excluded topics include improving logging practices without a security focus. After this process, we retain a set of 16 papers that mention security risks, patterns, or countermeasures in logging code. Two authors of this paper then independently review these papers and derive an initial set of codes that reflect recurring categories of logging-related security issues. In parallel, we manually inspect the filtered CWE entries and CVE reports to extract relevant vulnerability characteristics. The authors perform open coding and card sorting on the collected codes to group them into higher-level themes. Through iterative discussion, overlapping or ambiguous codes are merged or refined to ensure conceptual clarity and semantic consistency. This process results in a set of four categories and 10 detailed patterns that form our taxonomy of logging code security issues.

### 3.2 Investigation on Real-world Issues

In Stage 2, we investigate real-world security issues in logging code and examine whether the derived taxonomy can effectively categorize these issues. To this end, we collect issue reports from open-source project trackers, including Jira and GitHub. We first retrieve all logging-related issue reports by conducting a broad search using general logging keywords (e.g., "log", "logging", "sensitive information") with specific status (e.g., "Close", "Open", "In Progress", "Resolved", "Fixed") from 2019 to 2025. To maximize coverage, this initial search does not restrict the choice of projects and returns 949 issue reports across 93 projects. We then perform a filtering process to identify those relevant to security. Specifically, we remove low-quality or unsuitable reports (e.g., toy projects, incomplete descriptions, or duplicates) at the outset, and retain only those with clear descriptions, available solutions, and explicit evidence of being security-related. In particular, we only keep reports whose status is *"Resolved"*, *"Closed"*, or for which a solution is explicitly provided. This process yields our curated dataset, **SecLogging**, with 101 security-related logging issue reports from 63 projects.

With this list, we annotate each issue according to the taxonomy constructed in Stage 1. Specifically, two authors independently review the issue description, developer discussions, and corresponding code changes, and annotate the report with one of the taxonomy's categories and patterns. Disagreements are resolved through discussion until consensus is reached. Following prior studies [26, 49], we use Cohen's Kappa [43] to measure the agreement of results between the two authors. The Cohen's Kappa value in this process is 0.762, which indicates a substantial agreement. During this annotation process, we did not identify any new categories or patterns. In addition, for each report, we extract both the problematic code snippet (before the repair) and the corrected version (after the repair), thereby constructing a paired dataset that serves as the ground truth for subsequent experiments on detecting and repairing them with LLMs. Table 1 summarizes the distribution of the annotated issue reports across the taxonomy's categories and patterns. More details will be discussed in RQ1 of Section 5.

### 3.3 Framework for Detecting and Repairing Logging Code Security Issues with LLMs

To examine the capabilities of LLMs in detecting and repairing the logging code security issues, we propose a structured analysis framework that consists of three components: *Contextual Enrichment*, *Decomposed Task Execution*, and *Structured Output Generation*.

**Contextual Enrichment.** Code snippets alone are often insufficient for identifying security issues. We therefore enrich the input with additional contextual knowledge. Specifically, we provide two sources of knowledge: (1) **Security Pattern Explanation (+E)**, which describes the definition and characteristics of a security pattern, and (2) **Issue Description (+D)** taken from the original issue report, which provides human-written details of the problem, and is used to simulate scenarios where developers already possess some prior knowledge about the issue. These contexts can be supplied individually or in combination (+D+E) to enhance the LLMs' understanding.

**Decomposed Task Execution.** We decompose the overall analysis-and-repair process into three sub-tasks that mirror a human expert's workflow, including (1) **Detection:** determining whether a logging code security issue exists; (2) **Classification:** mapping the issue to a category and pattern in our taxonomy; and (3) **Remediation:** proposing a concrete repair suggestion.

**Structured Output Generation.** Finally, we constrain the results to a predefined, machine-parsable format, which is consistent in different experimental settings and easy for developers to understand systematically. The complete implementation details of our framework, including the corresponding prompt templates, can be found in our replication package [5].

## 4 Experimental Setup

In this section, we describe the design and implementation of our experiments to evaluate the effectiveness of LLMs in detecting and repairing security issues in logging code. We aim to answer the following research questions:

- **RQ1:** What security issues exist in logging code?
- **RQ2:** How effective are LLMs in detecting security issues in logging code?
- **RQ3:** To what extent can LLMs repair security issues in logging code?

### 4.1 Evaluation Metrics

The overall detection and repair process is formulated as a multi-step problem. Specifically, the LLM first examines whether the input logging code contains a security issue. If an issue is detected, it identifies the category of security issues as the parent class, along with the corresponding patterns as the sub-class. It then provides repaired code based on input logging code.

**Evaluation Metrics for Detection.** We analyze the output results which contain both category and pattern for each issue against the ground truth, calculating accuracy for the overall category, overall pattern, as well as individual accuracies for each category and pattern. Specifically, the accuracy is computed as:

$$\text{Accuracy}_{\text{Category/pattern}} = \frac{\text{Number of Correctly Detected Instances}}{\text{Total Instances of the Category / Pattern}}$$

**Evaluation Metrics for Repaired Code Patch.** For the evaluation of repaired code, we combine textual similarity with manual examination. For textual similarity between the solution provided by the LLM and ground truth, we adopt two commonly used measurements: Sequence Similarity and Jaccard Similarity. Sequence Similarity captures structural code changes and small-scale edits such as variable or function name modifications by measuring character-level differences. Meanwhile, Jaccard Similarity better reflects semantic similarity in terms of API usage and keywords by comparing vocabulary overlap, regardless of their exact ordering. These complementary metrics help assess different aspects of code similarity, and they are computed as follows:

$$\text{SequenceSim}(A, B) = 1 - \frac{\text{EditDistance}(A, B)}{max(|A|, |B|)}$$

$$\text{JaccardSim}(A, B) = \frac{|W_A \cap W_B|}{|W_A \cup W_B|}$$

where $W_A$, $W_B$ are the vocabulary sets of texts A and B, $W_A \cap W_B$ is the vocabulary intersection, and $W_A \cup W_B$ is the vocabulary union.

We then introduce an overall **Patch Similarity** as an average of the two metrics:

$$\text{PatchSim}(A, B) = \frac{\text{SequenceSim}(A, B) + \text{JaccardSim}(A, B)}{2}$$

This combined metric balances sensitivity to fine-grained structural edits (captured by Sequence Similarity) with robustness to semantic equivalence at the token level (captured by Jaccard Similarity). By averaging both, Patch Similarity provides a more comprehensive estimate of how closely an LLM-generated patch matches the ground truth, mitigating the biases of relying on only one type of similarity. In other words, a patch that differs in variable naming but preserves functional keywords should still be considered highly similar, while one that reuses many keywords but substantially changes the structure should be penalized. Apart from textual similarity, we also manually examine the effectiveness of repair suggestion that is elaborated in Section 7.

### 4.2 Studied LLMs and Implementation Details

In our experiments, we examine both open-source and closed-source LLMs. For open-source models, we select seven representative LLMs from three model families: Llama [22] (Llama3.3), Qwen series [47, 56] (Qwen2.5-72b, Qwen3-4b, Qwen3-32b), and DeepSeek [16, 17] (DeepSeek-R1-32b, DeepSeek-R1, DeepSeek-V3). For closed-source LLM, we select GPT-4.1-nano, primarily considering the affordability of API cost. For Llama3.3, Qwen series, and DeepSeek-R1-32b, we deploy their public checkpoints locally on our machine and invoke them using the Ollama framework [4]. For GPT-4.1-nano, DeepSeek-V3, and DeepSeek-R1, we utilize their official, publicly available APIs. All experiments are conducted on a Linux server running Ubuntu 20.04.6 LTS, equipped with an AMD 32-core processor, 1TB RAM, and eight NVIDIA A6000 GPUs.

Table 1. Taxonomy of Logging Security Issues with Instance Counts in *SecLogging*

| Security Issue Categories | Security Issue Patterns | # Instances (Category) | # Instances (Pattern) |
|---|---|---|---|
| **IL:** Insecure Log Storage & Access Control | **IL-At:** Risk of Log Injection Attacks<br>**IL-Pa:** Publicly Accessible Logs<br>**IL-Lv:** Insecure Logging Level Configuration | 13 | 2<br>1<br>10 |
| **SS:** Sensitive Information Exposure | **SS-Cr:** Credentials Leakage<br>**SS-Cf:** Configuration Data Exposure<br>**SS-Ur:** User Private Data Leakage | 38 | 20<br>7<br>11 |
| **RM:** Improper Redaction or Masking | **RM-Ms:** Missing Masking/Redaction<br>**RM-Ft:** Faulty Masking/Obfuscation | 43 | 36<br>7 |
| **EE:** Error & Exception Message Exposure | **EE-Ex:** Exception Leakage<br>**EE-St:** Stack Trace Leakage | 7 | 5<br>2 |
| | | *Total* | **101** |

## 5 Results

In this section, we discuss the results of our RQs.

### 5.1 RQ1: What security issues exist in logging code?

*5.1.1 Motivation.* In this RQ, we focus on what security issues exist in logging code, and how can they be systematically categorized. Through our comprehensive investigation (cf. Section 3), we derive a taxonomy consisting of four logging code security issue categories and 10 corresponding patterns, and annotate real-world issue reports using our taxonomy.

*5.1.2 Results.* Table 1 summarizes the taxonomy with instance counts in our constructed benchmark dataset, *SecLogging*. For clarity and readability, some of the code snippets presented in this section have been simplified to highlight only the parts relevant to the security problem. Where applicable, we also attach CWE/CVE entries that share similar characteristics with the pattern. Note that these may not represent a one-to-one match, as our patterns are abstracted from both prior literature and CWE/CVE descriptions.

**Insecure Log Storage and Access Control (IL).** This category refers to unsafe logging access and storage practices, where log files or output channels lack appropriate security protections. It emphasizes problems in how logs are stored or configured, rather than in the content being logged. This category has 13 instances in total and includes three patterns: IL-At (2 instances), IL-Pa (1 instances), and IL-Lv (10 instances).

- **IL-At** *Risk of Log Injection Attacks:* Log entries lack of filtering, enabling attackers to inject malicious content.
  **Real-world Example:** In Apache Karaf (KARAF-7061), the default Log4J2 configuration was vulnerable to log injection attacks. The issue originated from the use of the `log4j2.pattern` property with the "%m" directive, which prints raw log messages without escaping. As a result, attackers could inject carriage return/line feed (CRLF) characters or HTML tags to forge new log entries or launch Cross-Site Scripting (XSS) attacks when logs were rendered in web-based viewers or security audits.

```
// KARAF-7061
color.trace = cyan
log4j2.pattern = %d{ISO8601} | %-5p | %-16t | %-32c{1} | %X{bundle.id} - %X{bundle.name} - %X{bundle.version}
    | %m%n
```

```
log4j2.out.pattern = \u001b[90m%d{HH:mm:ss.SSS}\u001b[0m%highlight{%-5level} {FATAL=${color.fatal},
   ERROR=${color.error}, WARN=${color.warn}, INFO=${color.info}, DEBUG=${color.debug}, TRACE=${color.trace}}
   \u001b[90m[%t]\u001b[0m %msg%n%throwable
```

***Similar CWE/CVE:*** CWE-434 *"Unrestricted Upload of File with Dangerous Type"*, CVE-2021-44228, CVE-2020-17449.

- **IL-Pa** *Publicly Accessible Logs:* Log files or output channels are open to access, allowing any unauthorized user to directly view internal operational details.

  ***Real-world Example:*** In Apache Kafka (KAFKA-698), brokers could mistakenly expose uncommitted messages to consumers. The problem was caused by incorrect handling of the high watermark, which defines the offset boundary for committed messages. Because of this flaw, read requests could go beyond the boundary and allow consumers to access uncommitted data that might later be rolled back, and further lead to potential data leakage and inconsistency.

```scala
// KAFKA-698 (simplified)
val offsets =
  if (assignOffsets) {
    val firstOffset = nextOffset.get
    validMessages = validMessages.assignOffsets(nextOffset, messageSetInfo.codec)
    val lastOffset = nextOffset.get - 1
    (firstOffset, lastOffset)
  } else {
    if (!messageSetInfo.offsetsMonotonic)
      throw new IllegalArgumentException("Out of order offsets")
    nextOffset.set(messageSetInfo.lastOffset + 1)
    (messageSetInfo.firstOffset, messageSetInfo.lastOffset)
  }
```

***Similar CWE/CVE:*** CWE-312 *"Cleartext Storage of Sensitive Information"*, CVE-2023-6136.

- **IL-Lv** *Insecure Logging Level Configuration:* Improper log level settings, which result in sensitive data or system specifics being captured in logs.

  ***Real-world Example:*** In Apache Camel (CAMEL-14150), the "ExecCommand" class logged sensitive command details such as executables and arguments through its "ExecCommand" method. Because this information was recorded at the "INFO" level, internal system details were exposed in standard application logs.

```java
// CAMEL-14150 (simplified)
public class ExecCommand {
    private String executable;
    private List<String> args;
    private String workingDir;
    private File outFile;

    @Override
    public String toString() {
        return "ExecCommand [args=" + args + ", executable=" + executable
            + ", workingDir=" + workingDir + ", outFile=" + outFile + "]";
    }
}
```

***Similar CWE/CVE:*** CVE-2024-32474.

**Sensitive Information Exposure (SS).** This category covers cases where sensitive data is recorded directly or indirectly in logging code, leading to security or privacy leaks. This category includes three patterns (i.e., *SS-Cr, SS-Cf, SS-Ur*). This category has 38 instances in total and includes three patterns: SS-Cr (20 instances), SS-Cf (7 instances), and SS-Ur (11 instances).

- **SS-Cr** *Credentials Leakage:* Logs capture sensitive credentials like accounts, passwords, API keys, or OAuth tokens, which could be stolen and used in attacks.

  ***Real-world Example:*** In Apache Zeppelin (ZEPPELIN-2733), DEBUG-level logs printed sensitive cryptographic information, including an encrypted user key and its initialization vector (i.e.,

"*IV*"). An attacker with access to these logs could obtain both values, which allow them to analyze the encryption scheme or attempt offline decryption.

```
// ZEPPELIN-2733 (simplified)
private String getAuthKey(String userKey) {
    LOG.debug("Encrypted user key is {}", userKey);
    return decrypt(userKey, token.hashCode());
}

private String decrypt(String value, String initVector) {
    LOG.debug("IV is {}, IV length is {}", initVector, initVector.length());
    // ...
}
```

***Similar CWE/CVE:*** CWE-532 *"Insertion of Sensitive Information into Log File"*, CVE-2024-5908.

- **SS-Cf** *Configuration Data Exposure:* Log outputs include database connection URLs, critical file paths, environment variables, or other system configurations that attackers could exploit for infiltration.

  ***Real-world Example:*** In Apache Samza (SAMZA-589), the "MapConfig" class logged its entire configuration map via an unfiltered "toString()" method. As a result, sensitive values such as passwords, API keys, and database connection strings were exposed in plaintext in both logs and the application UI.

```
// SAMZA-589 (simplified)
public class MapConfig extends Config {
    private final Map<String, String> map;

    @Override
    public String toString() {
        return map.toString(); // Exposes full config including credentials
    }
}
```

***Similar CWE/CVE:*** CWE-532 *"Insertion of Sensitive Information into Log File"*, CVE-2024-32474.

- **SS-Ur** *User private data Leakage:* Log output includes user cookies, usernames, password, phone numbers, transaction records, or other personal data, leading to compliance issues and privacy violations.

  ***Real-world Example:*** In Apache Hadoop (HADOOP-17510), a trace-level logging statement in the "AuthCookieHandler" printed the contents of a user's authentication cookie. Even though logged at TRACE level, such cookies are sensitive and could be exploited for session hijacking if attackers gained access to the logs.

```
// HADOOP-17510 (simplified)
public class AuthCookieHandler {
    private HttpCookie authCookie;

    public void setAuthCookie(HttpCookie cookie, HttpCookie oldCookie) {
        this.authCookie = cookie;
        LOG.trace("Setting token value to {} ({})", authCookie, oldCookie);
    }
}
```

***Similar CWE/CVE:*** CWE-532 *"Insertion of Sensitive Information into Log File"*, CVE-2024-5557.

**Improper Redaction or Masking (RM).** This category covers insufficient redaction or masking in logging code, where sensitive information that should be hidden is either overlooked or handled poorly. This category highlights whether the logging code properly handles potentially risky information. This category includes two patterns (i.e., *RM-Ms, RM-Ft*). This category has 43 instances in total and includes three patterns: RM-Ms (36 instances), and RM-Ft (7 instances).

- **RM-Ms** *Missing Masking/Redaction:* Sensitive details are left entirely exposed, or dumping too much information that is hard to catch if sensitive information is leaked, exposing raw content directly in logs.
  *Real-world Example:* In Apache Hive (HIVE-9994), the query redaction mechanism is performed in "`redactQuery()`" after the query is being logged in "`PerfLogBegin()`". This ordering causes raw query strings containing confidential information (e.g., table names, user identifiers) to be logged before redaction, risking exposure of confidential details to anyone with access to logs.

```
// HIVE-9994 (simplified)
sem.validate();
perfLogger.PerfLogBegin(CLASS_NAME, PerfLogger.DRIVER_RUN, command);
plan = new QueryPlan(command, sem, perfLogger.getStartTime(), queryId, ...);
String queryStr = plan.getQueryStr();
for (Redactor r : getHooks(Redactor.class)) {
    queryStr = r.redactQuery(queryStr);
}
conf.setVar(HiveConf.ConfVars.HIVEQUERYSTRING, queryStr);
```

  *Similar CWE/CVE:* CWE-779 *"Logging of Excessive Data"*, CVE-2024-45784, CVE-2025-6711.

- **RM-Ft** *Faulty Masking/Obfuscation:* Even when masking is implemented, defects still allow some sensitive content to slip through in logging code.
  *Real-world Example:* In Apache Kafka (KAFKA-4056), the "`logUnused()`" method failed to apply masking when logging unrecognized configuration properties. As a result, if a sensitive key (e.g., a password) was misspelled, its plaintext value was written directly to the logs and bypasses the intended redaction mechanism.

```
// KAFKA-4056
public void logUnused() {
    for (String key : unused())
        log.warn("The configuration {} = {} was supplied but isn't a known config.", key,
    this.originals.get(key));
}
```

  *Similar CWE/CVE:* CVE-2025-53650, CVE-2024-32474.

**Error and Exception Message Exposure (EE).** This category covers the exposure of error or exception details in logging code, with messages or stack traces disclosing sensitive data or internal system workflows. This category specifically points to critical details leaking via exception handling routes like error messages or stack traces. This category has 7 instances in total and includes two patterns: EE-Ex (5 instances), and EE-St (2 instances).

- **EE-Ex** *Exception Leakage:* Exception message includes keys, file paths, config parameters, or other sensitive information, this allows attackers gain insights into the system's internals.
  *Real-world Example:* In Apache Hive (HIVE-20644), exception messages exposed sensitive runtime data by directly concatenating raw parameters, such as "`argumentString`" or "`rowString`", into error messages. When these exceptions were logged, user data and system details were leaked into log files.

```
// HIVE-20644 (simplified)
try {
    // Sensitive argumentString leaked in exception
    throw new HiveException(
        "Unable to execute method " + m + " with arguments " + argumentString, e);
} catch (Exception e) {
    // Sensitive rowString leaked in exception
    throw new HiveException(
        "Hive Runtime Error while processing row " + rowString, e);
}
```

Table 2. Accuracy (%) of LLMs on **SecLogging** (RQ2).

| Model | Setting | EE | IL | RM | SS | EE-Ex | EE-St | IL-At | IL-Lv | IL-Pa | RM-Ft | RM-Ms | SS-Cf | SS-Cr | SS-Ur | Category Average | Pattern Average |
|---|---|---|---|---|---|---|---|---|---|---|---|---|---|---|---|---|---|
| GPT-4.1-nano | Base | 14.3 | 20.0 | 10.0 | 87.2 | 0.0 | 0.0 | 100.0 | 0.0 | 0.0 | 0.0 | 11.1 | 14.3 | 70.0 | 0.0 | 41.6 | 20.8 |
|  | Base+E | 28.6 | 13.3 | 22.5 | 87.2 | 0.0 | 0.0 | 100.0 | 10.0 | 0.0 | 0.0 | 13.9 | 42.9 | 15.0 | 36.4 | 46.5 | 17.8 |
|  | Base+D | 28.6 | 13.3 | 12.5 | **92.3** | 20.0 | 0.0 | 50.0 | 0.0 | 0.0 | 0.0 | 13.9 | 28.6 | 55.0 | 0.0 | 44.5 | 19.8 |
|  | Base+E+D | 28.6 | 26.7 | 7.5 | 89.7 | 20.0 | 50.0 | 100.0 | 10.0 | 0.0 | 0.0 | 2.8 | 28.6 | 35.0 | 27.3 | 43.6 | 17.8 |
| Llama3.3 | Base | 14.3 | 13.3 | 17.5 | 71.8 | 20.0 | 0.0 | 100.0 | 0.0 | 0.0 | 0.0 | 19.4 | 0.0 | 75.0 | 18.2 | 37.6 | 26.7 |
|  | Base+E | 42.9 | 20.0 | 20.0 | 56.4 | 60.0 | 0.0 | 100.0 | 10.0 | 0.0 | 0.0 | 19.4 | 14.3 | 65.0 | 18.2 | 35.6 | 28.7 |
|  | Base+D | 28.6 | 13.3 | 25.0 | 76.9 | 20.0 | 0.0 | 100.0 | 0.0 | 0.0 | 0.0 | 22.2 | 0.0 | 70.0 | 9.1 | 43.6 | 25.7 |
|  | Base+E+D | 57.1 | 13.3 | 20.0 | 53.8 | 40.0 | 0.0 | 100.0 | 0.0 | 0.0 | 0.0 | 19.4 | 0.0 | 55.0 | 9.1 | 34.7 | 22.8 |
| DeepSeek-R1-32b | Base | 0.0 | 0.0 | 2.5 | 30.8 | 0.0 | 0.0 | 0.0 | 0.0 | 0.0 | 0.0 | 3.0 | 0.0 | 30.0 | 0.0 | 12.9 | 6.9 |
|  | Base+E | 14.3 | 6.7 | 5.0 | 35.9 | 0.0 | 0.0 | 0.0 | 0.0 | 0.0 | 0.0 | 0.0 | 14.0 | 20.0 | 0.0 | 18.0 | 5.0 |
|  | Base+D | 14.3 | 0.0 | 2.5 | 30.8 | 20.0 | 0.0 | 0.0 | 0.0 | 0.0 | 0.0 | 2.8 | 0.0 | 40.0 | 0.0 | 13.9 | 9.9 |
|  | Base+E+D | 28.6 | 6.7 | 5.0 | 28.2 | 20.0 | 50.0 | 50.0 | 0.0 | 0.0 | 0.0 | 2.8 | 14.3 | 15.0 | 9.1 | 15.8 | 8.9 |
| DeepSeek-R1 | Base | 28.6 | 13.3 | 12.5 | 79.5 | 40.0 | 0.0 | 50.0 | 0.0 | 0.0 | 0.0 | 8.3 | 28.6 | 60.0 | 27.3 | 39.6 | 22.8 |
|  | Base+E | 57.1 | 13.3 | 15.0 | 59.0 | 80.0 | 0.0 | 100.0 | 0.0 | 0.0 | 0.0 | 8.3 | 14.3 | 40.0 | 27.3 | 34.7 | 20.8 |
|  | Base+D | 57.1 | 27.0 | 27.5 | 71.8 | 60.0 | 0.0 | 50.0 | 20.0 | 0.0 | 0.0 | 16.7 | 0.0 | 60.0 | 36.4 | 48.5 | 29.7 |
|  | Base+E+D | 71.4 | 13.3 | **30.0** | 61.5 | 60.0 | 50.0 | 100.0 | 0.0 | 0.0 | 0.0 | 16.7 | 14.3 | 45.0 | 27.3 | 48.5 | 24.8 |
| DeepSeek-V3 | Base | 71.4 | 20.0 | 7.5 | 79.5 | 60.0 | 0.0 | 100.0 | 0.0 | 0.0 | 0.0 | 5.6 | 28.6 | 45.0 | 27.3 | 41.6 | 20.8 |
|  | Base+E | 71.4 | **33.3** | 12.5 | 76.9 | 40.0 | 0.0 | 100.0 | 20.0 | 0.0 | 0.0 | 5.6 | 42.9 | 40.0 | 27.3 | 44.6 | 22.8 |
|  | Base+D | **85.7** | 13.3 | 27.5 | 87.2 | 100.0 | 50.0 | 100.0 | 0.0 | 0.0 | 14.3 | 19.4 | 28.6 | 75.0 | 27.3 | **52.5** | **35.6** |
|  | Base+E+D | **85.7** | **33.3** | 20.0 | 76.9 | 80.0 | 50.0 | 100.0 | 20.0 | 0.0 | 14.3 | 11.1 | 28.6 | 55.0 | 27.3 | 48.5 | 29.7 |
| Qwen2.5-72B | Base | 0.0 | 13.3 | 5.0 | 74.4 | 0.0 | 0.0 | 100.0 | 0.0 | 0.0 | 0.0 | 2.8 | 28.6 | 70.0 | 9.1 | 32.7 | 19.8 |
|  | Base+E | 28.6 | 20.0 | 22.5 | 59.0 | 40.0 | 0.0 | 100.0 | 10.0 | 0.0 | 0.0 | 13.9 | 0.0 | 45.0 | 18.2 | 34.7 | 20.8 |
|  | Base+D | 28.6 | 13.3 | 5.0 | 74.4 | 0.0 | 0.0 | 100.0 | 0.0 | 0.0 | 0.0 | 5.6 | 0.0 | 70.0 | 0.0 | 34.7 | 17.8 |
|  | Base+E+D | 42.9 | 20.0 | 10.0 | 64.1 | 20.0 | 0.0 | 100.0 | 10.0 | 0.0 | 0.0 | 8.3 | 0.0 | 40.0 | 18.2 | 34.7 | 16.3 |
| Qwen3-4B | Base | 0.0 | 6.7 | 5.0 | 59.0 | 0.0 | 0.0 | 50.0 | 0.0 | 0.0 | 0.0 | 5.6 | 0.0 | 50.0 | 27.3 | 25.7 | 16.8 |
|  | Base+E | 14.3 | 20.0 | 17.5 | 48.7 | 20.0 | 0.0 | 50.0 | 10.0 | 0.0 | 0.0 | 16.7 | 42.9 | 40.0 | 9.1 | 29.7 | 20.8 |
|  | Base+D | 14.3 | 13.3 | 7.5 | 61.5 | 40.0 | 0.0 | 100.0 | 0.0 | 0.0 | 0.0 | 11.1 | 14.3 | 40.0 | 9.1 | 29.7 | 17.8 |
|  | Base+E+D | 14.3 | 27.0 | 7.5 | 43.6 | 20.0 | 0.0 | 100.0 | 20.0 | 0.0 | 0.0 | 8.3 | 14.3 | 40.0 | 9.1 | 24.8 | 17.8 |
| Qwen3-32B | Base | 0.0 | 20.0 | 25.0 | 61.5 | 0.0 | 0.0 | 100.0 | 10.0 | 0.0 | 14.3 | 25.0 | 28.6 | 65.0 | 0.0 | 36.6 | 27.7 |
|  | Base+E | 42.9 | 26.7 | 20.0 | 56.4 | 40.0 | 0.0 | 100.0 | 20.0 | 0.0 | 0.0 | 11.1 | 28.6 | 40.0 | 9.1 | 36.6 | 20.8 |
|  | Base+D | 42.9 | 20.0 | 20.0 | 51.3 | 0.0 | 50.0 | 100.0 | 10.0 | 0.0 | 0.0 | 19.4 | 28.6 | 55.0 | 0.0 | 33.7 | 23.8 |
|  | Base+E+D | 42.9 | 20.0 | 15.0 | 48.7 | 40.0 | 50.0 | 100.0 | 10.0 | 0.0 | 0.0 | 8.3 | 14.3 | 55.0 | 18.2 | 30.7 | 22.8 |

**Similar CWE/CVE:** CWE-209 *"Generation of Error Message Containing Sensitive Information"*, CVE-2019-16768.

- **EE-St** *Stack Trace Leakage:* Stack Traces logged reveal function call chains, internal class names, or library versions, this provides attackers with valuable intelligence on the system's architecture. **Real-world Example:** In Apache Commons Net (NET-618), when a "ParseException" is caught, the code directly calls "e.printStackTrace()". This method prints the full stack trace to the standard error stream, which is often captured in application logs. This exposed detailed internal information, including method call chains, class names, and line numbers, which attackers could exploit to map the system's architecture or identify vulnerable libraries.

```
// NET-618
try {
    file.setTimestamp(super.parseTimestamp(datestr));
} catch (ParseException e) {
    e.printStackTrace();
```

**Similar CWE/CVE:** CWE-117 *"Improper Output Neutralization for Logs"*, CVE-2020-4085.

> **RQ1 Summary:** We derive a taxonomy consisting of four categories and 10 patterns of logging code security issues, and observe that these issues are widely present in real-world projects. Among them, Improper Redaction or Masking (RM) is the most prevalent category with 43 instances in our collected issue reports.

## 5.2 RQ2: How effective are LLMs in detecting security issues in logging code?

***Motivation.*** LLMs have shown promising results in detecting general software security issues such as vulnerabilities [57]. However, it remains unclear whether these capabilities extend effectively to the domain of logging code, where security issues have distinct characteristics. Therefore, in this RQ, we investigate how efficiently LLMs can detect security issues in logging code.

***Approach.*** We run experiments on ***SecLogging*** and use the framework discussed in Section 3 to evaluate the effectiveness of LLMs in detecting security issues in logging code. For each contextual configuration (i.e., Base, Base*+D*, Base*+E*, Base*+D+E*), we compute the detection accuracy at both the category and pattern levels.

***Results.*** Table 2 presents the results of this RQ. **Adding an issue description (+D) generally provides the most consistent improvement across categories.** For instance, DeepSeek-R1's average category accuracy rises from 39.6% to 46.5% with this addition. Explanations (+E) also help in some cases, but their effect is less stable, and combining both description and explanation (+D+E) does not always yield better outcomes. In fact, richer context sometimes may lead to degraded results, such as Llama3.3, whose average accuracy drops to 34.7% under (+D+E). These findings suggest that concise, targeted guidance is often more effective, and when developers already possess some understanding of the issue (e.g., through an issue description), this knowledge can better guide LLMs to improve detection.

**LLM's capability plays an important role.** In particular, DeepSeek-V3 achieves the highest accuracy overall, reaching 85.7% on the EE category with (+D). Other LLMs like DeepSeek-R1 and Llama3.3 also perform competitively, while smaller models such as DeepSeek-R1-32b struggle compared to other larger LLMs, rarely exceeding 20% in category accuracy across settings.

**Performance varies substantially across categories and patterns.** Among categories, SS (Sensitive Information Exposure) is relatively easier to detect, with several LLMs exceeding 70% accuracy, while RM (Improper Redaction or Masking) remains the most difficult, often falling below 30%. Within patterns, concrete issues such as IL-At (Log Injection Attacks) and SS-Cr (Credentials Leakage) are reliably identified, with models frequently achieving above 70%. In contrast, abstract and logic-dependent patterns like RM-Ms (Missing Masking/Redaction) are more challenging, with almost all the LLMs performing poorly. It suggests that LLMs may handle well-defined, token-based flaws more effectively than subtle issues requiring reasoning about missing mechanisms.

> **RQ2 Summary:** Overall, the detection accuracy of LLMs at the category level varies widely, ranging from 12.9% to 52.5% on average across models and settings. Adding an issue description (+D) generally provides the most consistent improvement. In addition, performance differs considerably across patterns: models achieve high accuracy on concrete issues such as Sensitive Information Exposure (SS), but consistently struggle with abstract, logic-based flaws like Improper Redaction or Masking (RM), where accuracy often falls below 20%.

## 5.3 RQ3: To what extent can LLMs repair security issues in logging code?

Table 3. Patch Similarity for Different LLMs.

| Setting | GPT-4.1-nano | Llama3.3 | DeepSeek-R1-32b | DeepSeek-R1 | DeepSeek-V3 | Qwen2.5-72B | Qwen3-4B | Qwen3-32B | Avg. |
|---|---|---|---|---|---|---|---|---|---|
| Base | 29.8 | 43.4 | 24.2 | 60.6 | **69.3** | 48.5 | 36.6 | 40.5 | 44.1 |
| Base+E | 28.5 | 38.5 | 23.6 | 49.1 | 66.0 | 43.6 | 31.2 | 37.2 | 39.7 |
| Base+D | 27.2 | 34.1 | **18.8** | 58.5 | 68.3 | 42.9 | 31.4 | 35.0 | 39.5 |
| Base+E+D | 28.3 | 34.8 | 20.3 | 47.3 | 63.4 | 34.7 | 30.7 | 32.5 | 36.5 |
| *Average* | 28.4 | 37.7 | 21.7 | 53.9 | 66.8 | 42.4 | 32.5 | 36.3 | 40.0 |

*Motivation.* Recent advances show that LLMs can effectively support automated program repair by producing meaningful patches for buggy code [62]. This opens the question of whether such capabilities can be leveraged to handle security issues in logging code, which often involve domain-specific risks and knowledge. In this RQ, we examine how well LLMs can generate repairs for security issues in logging code.

*Approach.* We use the same set of LLMs and prompt configurations as in RQ2 and run experiments on **SecLogging**. The models are tasked with generating a repaired code patch for the identified security issue. We evaluate the results in two aspects: automated textual patch similarity, and manual evaluation.

❶ **Textual Patch Similarity Evaluation.** We compare generated patches against ground-truth repairs using the *Patch Similarity* discussed in Section 3.

❷ **Manual Evaluation.** To validate automated metrics and gain qualitative insights, two authors independently evaluate each case by examining the generated code patch. If the patch is considered to be reasonable and effective, the result is marked as *effective*. In cases of disagreement, the two authors discuss together to reach a consensus.

*Results.* We discuss the results of textual patch similarity and manual evaluation, respectively.

❶ **Textual Patch Similarity Evaluation.** Table 3 presents the similarity-based evaluation results. Different from detection accuracy, **adding contextual information does not consistently improve repair patch generation**. When comparing the four prompt settings, the Base setting achieves the highest overall average similarity (44.1%), while adding contextual information (i.e., +E, +D, or +E+D) generally reduces performance, with the lowest average observed for Base+E+D (36.5%). For example, DeepSeek-V3 peaks at 69.3% in the Base setting, while its performance drops to 63.4% under (+E+D). Similarly, Llama3.3 falls to 34.8% with (+E+D) from 43.4% in the setting of (Base). These results suggest that supplementary context can sometimes be helpful, but it may also introduce noise that reduces the ability to repair issues.

The results show **disparities in the similarity across different LLMs**. Overall, DeepSeek-V3 outperforms other LLMs, with Patch Similarity consistently above 63.4% and peaking at 69.3%. DeepSeek-R1 also performs strongly, reaching up to 60.6%. Other LLMs such as Qwen2.5-72B also achieve reasonable results (up to 48.5%), while smaller models like GPT-4.1-nano consistently remain below 30%.

❷ **Manual Evaluation.** Overall, **the results of our manual evaluation align with the Patch Similarity results.** Specifically, we examine repair results from the experiment of highest Patch Similarity (i.e., DeepSeek-V3 with Base) and the lowest (i.e., DeepSeek-R1-32b with +D). We find that DeepSeek-V3 attempt to repair all the issues in the dataset, and 84 of these repairs (83.2%) are labeled as effective. By contrast, DeepSeek-R1-32b provide repair suggestions for only 36 issues, of which 27 (75.0%) are effective. In addition, we also conduct case studies to analyze the results of high and low Patch Similarity in detail.

*Case Study.* Figure 3 shows a case where the LLM provides a repair with a relative high Patch Similarity (92.8%) with the developer's repair. The code is a Python "`__repr__`" method that exposes multiple fields of Personally Identifiable Information (PII), including the customer's name, address, and "pesel" (a national identification number), directly to the logs. This corresponds to the SS-Ur (User Private Data Leakage) pattern. Both the developer and the LLM remediate the issue by redacting the sensitive fields. The developer apply full masking, replacing values with asterisks, whereas the LLM use a more context-aware approach: general fields are fully masked, but the "pesel" number is partially redacted to reveal only the last four digits. This reflects a common industry practice that balances security with usability, which shows that the LLM can align with conventions for handling sensitive data.

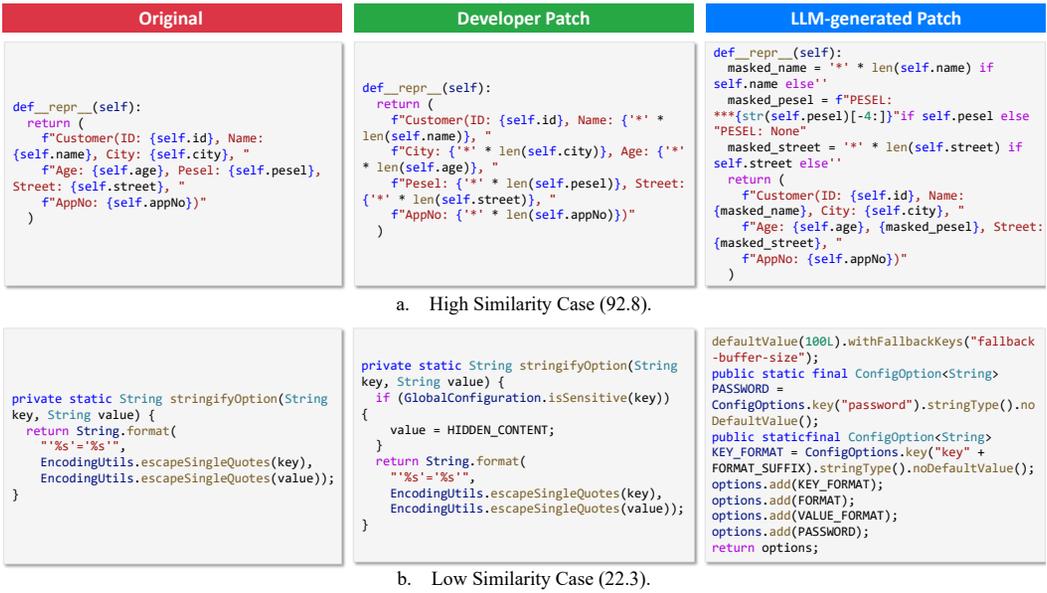

Fig. 3. Case study with different Patch Similarity.

Figure 3 also illustrates another case with a low patch similarity score (22.3%), which shows a tendency for LLMs to propose overly complex and indirect solutions. The issue in the original code exists in the `stringifyOption` method, which is used to format configuration options for Table API exception messages. The method performed no sensitivity checks, causing sensitive values, such as passwords, to be leaked in plaintext through simple string formatting. The developer's ground-truth repair is direct and effective. It modifies the `stringifyOption` method to integrate a `GlobalConfiguration.isSensitive(key)` check, which replaces any sensitive value with a `HIDDEN_CONTENT` placeholder before it is formatted into the final string.

In contrast, the LLM's output does not perform this simple, targeted modification. Instead, it suggests an unnecessary refactoring by introducing a new `ConfigOption` class to manage configuration variables. While this alternative approach might implicitly mask some values, it fails to directly address the insecure formatting in the original method, leaving a potential risk of data leakage. This change to the existing code logic greatly increases the complexity and cost of manual review and future maintenance.

> **RQ3 Summary:** Our automated and manual evaluation show that LLMs can provide repairs for logging code security issues with varying effectiveness. DeepSeek-V3 achieves the highest Patch Similarity (up to 69.3%), while smaller LLMs such as GPT-4.1-nano consistently underperform. Different from detection, adding contextual information does not help in the capability of providing repairs and can sometimes considerably reduce it.

## 6 Discussion
### 6.1 Analysis of Token Consumption

To better understand the trade-off between cost and effectiveness, we analyze the impact of input token usage across different prompt settings. Figure 4 illustrates that adding more context increases input tokens, but may or may not lead to performance improvement. For detection accuracy, we find that adding an issue description (+D) improves the accuracy from 36.5% to 37.4%, and the input token consumption increased from 3,003 to 3,688 on average. On the other hand, prompt with the most detailed contextual information (i.e., +E+D) appears to be among the least effective while also incurring the highest cost. This observation is more obvious for code fixing issues, where the Base prompt yields the highest Patch Similarity on average and +E+D achieves the worst results. These results suggest that for both detection and repair, concise and targeted prompts may provide a more favorable balance between cost and effectiveness, while overly detailed prompts may increase cost without improving outcomes.

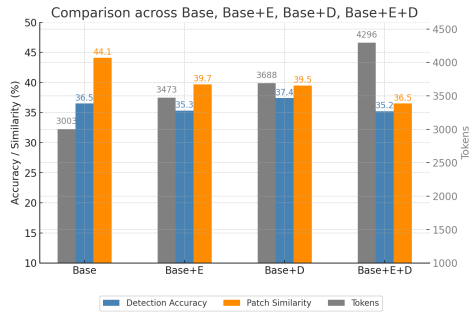

Fig. 4. Comparison of average detection accuracy and patch similarity across four settings' token usage.

### 6.2 Implications

We discuss the implications of our study for practitioners and researchers, respectively. Specifically, IMP-P# are implications for practitioners, and IMP-R# are for researchers.

**IMP-P1. Guidelines for secure logging and support for rapid incident response.** Our taxonomy of logging security issues provides a concrete guideline to help developers and engineers prevent security issues. It can also be codified into automated tools, such as custom security-focused linting rules. When a security incident occurs, the taxonomy can also be used a structured framework to accelerate investigation and response, potentially lowering Mean Time to Detect (MTTD) and Mean Time to Repair (MTTR). When using LLMs for security analysis, our results show that providing concise, specific context (e.g., an issue description) is crucial for improving detection accuracy.

**IMP-P2. Contextual information is not "the more, the better".** Our results show that concise, specific context can substantially improve detection accuracy (e.g., +D) . However, it is not always "the more, the better". For example, combining both explanation and issue description (i.e., +E+D ) degrades the performance of detection and repair, and increases the tokens used. This aligns with our token consumption analysis, which shows that more verbose prompts increase computational cost, but may also distract the LLM in analyzing the task. In other words, while targeted context improves effectiveness, excessive or redundant context may introduce unnecessary tokens without clear benefits. For practitioners, this highlights the importance of providing precise and minimal information to maximize detection accuracy while keeping costs manageable.

**IMP-P3. Human oversight in auditing and repair.** Practitioners should prioritize auditing "missing-type" issues that LLMs struggle to detect, such as mssing masking or redaction (*RM-Ms*). These represent critical blind spots for LLMs but are areas where human expertise is particularly valuable. In addition, caution is warranted when adopting LLM-generated fixes. Our study shows that additional context often reduces repair quality, and case studies reveal that LLMs may propose overly complex or even incorrect repair suggestions. Rigorous human review remains essential before integrating AI-suggested patches.

**IMP-R1. Benchmark for improving LLMs in secure logging.** Our results show that the current performance of LLMs in detecting and repairing logging code security issues is still limited and cannot yet be considered highly effective. The taxonomy and curated dataset introduced in this study can therefore serve as valuable benchmarks for evaluating and improving LLMs. By utilizing our benchmarks, researchers can systematically evaluate LLMs on real-world logging security issues and identify both their effective cases and their limitations. These evaluations provide a concrete basis for improving future LLMs and for proposing more reliable support for secure logging practices.

**IMP-R2. More cost-efficient and effective contextual enrichment strategy.** Our results demonstrate the importance of advancing both prompting strategies and the LLMs. On the one hand, there is a clear need for cost-efficient and performance-oriented prompting and contextual enrichment approaches. On the other hand, current LLMs often struggle when presented with complex or redundant contextual information. Improving LLMs that can better process and prioritize heterogeneous inputs, rather than being distracted by them, will be essential for enabling more robust detection and repair of logging security issues.

## 7 Threats to Validity

**Internal Validity.** Our manual examination of repairs may introduce subjectivity and human error. However, our results achieved a Cohen's kappa value of 0.728, indicating substantial inter-rater agreement and consistency in our assessments. The selection and annotation of logging code samples may also be influenced by our prior knowledge and assumptions. Furthermore, the experimental setup, including prompt design and evaluation criteria, could affect the outcomes. We mitigated these risks through multiple rounds of validation, consensus discussions, and by complementing automated metrics with manual review.

**External Validity.** In this study, we evaluate a set of LLMs from different families and settings, including both open-source and closed-source models. While this selection covers a range of current capabilities, it may not reflect the full diversity of available or future LLMs. However, by including diverse models and configurations, our results provide a broadly generalizable view of current LLM capabilities for logging code security. We collect logging code samples from real-world issue reports on platforms such as Jira and GitHub. Although these sources provide practical and relevant examples, they may not encompass all possible logging practices, programming languages, or security issues found in other environments. As a result, our benchmark may not fully represent the broader landscape of logging code security. Future research should expand the dataset to include additional sources to further enhance generalizability.

**Construct Validity.** A potential threat is that our automated evaluations primarily rely on textual similarity metrics, which may not fully capture the functional correctness of the generated repair patches. Textually similar code may not imply functional equivalence. To mitigate this, we complement automated similarity metrics with manual evaluation, where we assessed the functional correctness of the repaired code. This dual approach helps ensure our results reflect both syntactic and semantic aspects of code repair. Future work could further strengthen construct validity by incorporating automated semantic analysis or dynamic testing.

## 8 Conclusion

In this paper, we investigate logging code security through a taxonomy of four categories and 10 scenario patterns, and we construct the ***SecLogging*** benchmark from real-world issue reports.

Our experiments show that current LLMs generally face limitations in detecting and repairing logging code security issues. However, we find concise issue descriptions can improve detection accuracy, while over verbose prompts often increase token cost without providing additional benefits. Our findings highlight the importance of deriving more cost-efficient prompting strategies and advancing LLMs with stronger capabilities to analyze logging code security issues.

### Data Availability

Our replication package is available and can be accessed using the link [5].

### References

[1] 2025. CVE: Common Vulnerabilities and Exposures. https://www.cve.org/. Last accessed September 2025.
[2] 2025. Log4Shell. https://en.wikipedia.org/wiki/Log4Shell. Last accessed September 2025.
[3] 2025. MITRE - CWE List. https://cwe.mitre.org/data/index.html. Last accessed September 2025.
[4] 2025. Ollama. https://ollama.com. Last accessed September 2025.
[5] 2026. Replicaton Package. https://github.com/defects4log/SecLogging. Last accessed April 2026.
[6] Roozbeh Aghili, Heng Li, and Foutse Khomh. 2025. Protecting Privacy in Software Logs: What Should Be Anonymized? *Proceedings of the ACM on Software Engineering* 2, FSE (2025), 1317–1338.
[7] Roozbeh Aghili, Xingfang Wu, Foutse Khomh, and Heng Li. 2025. SDLog: A Deep Learning Framework for Detecting Sensitive Information in Software Logs. *arXiv preprint arXiv:2505.14976* (2025).
[8] Adil Ahmad, Sangho Lee, and Marcus Peinado. 2022. Hardlog: Practical tamper-proof system auditing using a novel audit device. In *2022 IEEE Symposium on Security and Privacy (SP)*. IEEE, 1791–1807.
[9] Tuan-Dung Bui, Thanh Trong Vu, Thu-Trang Nguyen, Son Nguyen, and Hieu Dinh Vo. 2025. Correctness Assessment of Code Generated by Large Language Models Using Internal Representations. *arXiv preprint arXiv:2501.12934* (2025).
[10] Jeanderson Cândido, Maurício Aniche, and Arie Van Deursen. 2021. Log-based software monitoring: a systematic mapping study. *PeerJ Computer Science* 7 (2021), e489.
[11] Boyuan Chen and Zhen Ming Jiang. 2017. Characterizing and Detecting Anti-Patterns in the Logging Code. In *2017 IEEE/ACM 39th International Conference on Software Engineering (ICSE)*.
[12] Haoyu Chen, Shanshan Tu, Chunye Zhao, and Yongfeng Huang. 2016. Provenance cloud security auditing system based on log analysis. In *2016 IEEE International Conference of Online Analysis and Computing Science (ICOACS)*. IEEE, 155–159.
[13] Junkai Chen, Xing Hu, Zhenhao Li, Cuiyun Gao, Xin Xia, and David Lo. 2024. Code search is all you need? improving code suggestions with code search. In *Proceedings of the IEEE/ACM 46th International Conference on Software Engineering (ICSE)*. 1–13.
[14] Junkai Chen, Huihui Huang, Yunbo Lyu, Junwen An, Jieke Shi, Chengran Yang, Ting Zhang, Haoye Tian, Yikun Li, Zhenhao Li, Xin Zhou, Xing Hu, and David Lo. 2025. SecureAgentBench: Benchmarking Secure Code Generation under Realistic Vulnerability Scenarios. *arXiv preprint arXiv:2509.22097* (2025).
[15] Junkai Chen, Zhenhao Li, Qiheng Mao, Xing Hu, Kui Liu, and Xin Xia. 2025. Understanding practitioners' expectations on clear code review comments. *Proceedings of the ACM on Software Engineering* 2, ISSTA (2025), 1257–1279.
[16] DeepSeek-AI, Daya Guo, Dejian Yang, and et al. 2025. DeepSeek-R1: Incentivizing Reasoning Capability in LLMs via Reinforcement Learning.
[17] DeepSeek-AI, Aixin Liu, Bei Feng, Bing Xue, and et al. Wang. 2025. DeepSeek-V3 Technical Report. doi:10.48550/arXiv.2412.19437
[18] Zishuo Ding, Yiming Tang, Yang Li, Heng Li, and Weiyi Shang. 2023. On the temporal relations between logging and code. In *2023 IEEE/ACM 45th International Conference on Software Engineering (ICSE)*. IEEE, 843–854.
[19] Zishuo Ding, Yiming Tang, Yang Li, Heng Li, and Weiyi Shang. 2023. On the Temporal Relations between Logging and Code. In *2023 IEEE/ACM 45th International Conference on Software Engineering (ICSE)*.
[20] Jiahao Fan, Yi Li, Shaohua Wang, and Tien N Nguyen. 2020. AC/C++ code vulnerability dataset with code changes and CVE summaries. In *Proceedings of the 17th international conference on mining software repositories*. 508–512.
[21] Qiang Fu, Jieming Zhu, Wenlu Hu, Jian-Guang Lou, Rui Ding, Qingwei Lin, Dongmei Zhang, and Tao Xie. 2014. Where do developers log? an empirical study on logging practices in industry. In *Companion Proceedings of the 36th International Conference on Software Engineering*. 24–33.
[22] Aaron Grattafiori, Abhimanyu Dubey, Abhinav Jauhri, Abhinav Pandey, and et al. 2024. The Llama 3 Herd of Models.
[23] Mehran Hassani, Weiyi Shang, Emad Shihab, and Nikolaos Tsantalis. 2018. Studying and detecting log-related issues. *Empirical Softw. Engg.* (2018).

[24] Yi Wen Heng, Zeyang Ma, Zhenhao Li, Dong Jae Kim, et al. 2024. Studying and Benchmarking Large Language Models For Log Level Suggestion. *arXiv preprint arXiv:2410.08499* (2024).
[25] Xing Hu, Feifei Niu, Junkai Chen, Xin Zhou, Junwei Zhang, Junda He, Xin Xia, and David Lo. 2025. Assessing and Advancing Benchmarks for Evaluating Large Language Models in Software Engineering Tasks. *arXiv preprint arXiv:2505.08903* (2025).
[26] Xing Hu, Xin Xia, David Lo, Zhiyuan Wan, Qiuyuan Chen, and Thomas Zimmermann. 2022. Practitioners' Expectations on Automated Code Comment Generation. In *Proceedings of the 44th international conference on software engineering*. 1693–1705.
[27] Zhihan Jiang, Jinyang Liu, Zhuangbin Chen, Yichen Li, Junjie Huang, Yintong Huo, Pinjia He, Jiazhen Gu, and Michael R Lyu. 2024. Lilac: Log parsing using llms with adaptive parsing cache. *Proceedings of the ACM on Software Engineering* 1, FSE (2024), 137–160.
[28] Zhihan Jiang, Jinyang Liu, Junjie Huang, Yichen Li, Yintong Huo, Jiazhen Gu, Zhuangbin Chen, Jieming Zhu, and Michael R Lyu. 2024. A large-scale evaluation for log parsing techniques: How far are we?. In *Proceedings of the 33rd ACM SIGSOFT International Symposium on Software Testing and Analysis*. 223–234.
[29] Staffs Keele et al. 2007. *Guidelines for performing systematic literature reviews in software engineering*. Technical Report. Technical report, ver. 2.3 ebse technical report. ebse.
[30] Amirmahdi Khosravi Tabrizi, Naser Ezzati-Jivan, and Francois Tetreault. 2024. An Adaptive Logging System (ALS): Enhancing Software Logging with Reinforcement Learning Techniques. In *Proceedings of the 15th ACM/SPEC International Conference on Performance Engineering*. 37–47.
[31] Heng Li, Weiyi Shang, Bram Adams, Mohammed Sayagh, and Ahmed E Hassan. 2020. A qualitative study of the benefits and costs of logging from developers' perspectives. *IEEE Transactions on Software Engineering* 47, 12 (2020), 2858–2873.
[32] Yichen Li, Yintong Huo, Renyi Zhong, Zhihan Jiang, Jinyang Liu, Junjie Huang, Jiazhen Gu, Pinjia He, and Michael R. Lyu. 2024. Go Static: Contextualized Logging Statement Generation. *Proc. ACM Softw. Eng.* (2024).
[33] Yi Li, Shaohua Wang, and Tien N Nguyen. 2021. Vulnerability detection with fine-grained interpretations. In *Proceedings of the 29th ACM Joint Meeting on European Software Engineering Conference and Symposium on the Foundations of Software Engineering*. 292–303.
[34] Zhenhao Li, An Ran Chen, Xing Hu, Xin Xia, Tse-Hsun Chen, and Weiyi Shang. 2023. Are they all good? studying practitioners' expectations on the readability of log messages. In *2023 38th IEEE/ACM International Conference on Automated Software Engineering (ASE)*. IEEE, 129–140.
[35] Zhenhao Li, Tse-Hsun Chen, and Weiyi Shang. 2020. Where Shall We Log? Studying and Suggesting Logging Locations in Code Blocks. In *35th IEEE/ACM International Conference on Automated Software Engineering, ASE 2020*. 361–372.
[36] Zhenhao Li, Tse-Hsun Chen, Jinqiu Yang, and Weiyi Shang. 2019. DLFinder: Characterizing and Detecting Duplicate Logging Code Smells. In *2019 IEEE/ACM 41st International Conference on Software Engineering (ICSE)*.
[37] Zhenhao Li, Heng Li, Tse-Hsun Chen, and Weiyi Shang. 2021. Deeplv: Suggesting log levels using ordinal based neural networks. In *2021 IEEE/ACM 43rd International Conference on Software Engineering (ICSE)*. IEEE, 1461–1472.
[38] Zhenhao Li, Chuan Luo, Tse-Hsun Chen, Weiyi Shang, Shilin He, Qingwei Lin, and Dongmei Zhang. 2023. Did we miss something important? Studying and exploring variable-aware log abstraction. In *2023 IEEE/ACM 45th International Conference on Software Engineering (ICSE)*. 830–842.
[39] Lipeng Ma, Weidong Yang, Bo Xu, Sihang Jiang, Ben Fei, Jiaqing Liang, Mingjie Zhou, and Yanghua Xiao. 2024. Knowlog: Knowledge enhanced pre-trained language model for log understanding. In *Proceedings of the 46th ieee/acm international conference on software engineering*. 1–13.
[40] Zeyang Ma, Dong Jae Kim, and Tse-Hsun Chen. 2024. LibreLog: Accurate and Efficient Unsupervised Log Parsing Using Open-Source Large Language Models. *arXiv preprint arXiv:2408.01585* (2024).
[41] Qiheng Mao, Zhenhao Li, Xing Hu, Kui Liu, Xin Xia, and Jianling Sun. 2025. Towards explainable vulnerability detection with large language models. *IEEE Transactions on Software Engineering* (2025).
[42] Antonio Mastropaolo, Luca Pascarella, and Gabriele Bavota. 2022. Using deep learning to generate complete log statements. In *Proceedings of the 44th international conference on software engineering*. 2279–2290.
[43] Mary L. McHugh. 2012. Interrater reliability: the kappa statistic. *Biochemia Medica* 22, 3 (2012), 276–282.
[44] Andriy Miranskyy, Abdelwahab Hamou-Lhadj, Enzo Cialini, and Alf Larsson. 2016. Operational-log analysis for big data systems: Challenges and solutions. *IEEE Software* 33, 2 (2016), 52–59.
[45] Zulie Pan, Yu Chen, Yuanchao Chen, Yi Shen, and Yang Li. 2022. LogInjector: Detecting web application log injection vulnerabilities. *Applied Sciences* 12, 15 (2022), 7681.
[46] Yun Peng, Jun Wan, Yichen Li, and Xiaoxue Ren. 2025. Coffe: A code efficiency benchmark for code generation. *Proceedings of the ACM on Software Engineering* 2, FSE (2025), 242–265.
[47] Qwen, An Yang, Baosong Yang, Beichen Zhang, and et al. 2025. Qwen2.5 Technical Report.

[48] Xiaoxue Ren, Jun Wan, Yun Peng, Zhongxin Liu, Ming Liang, Dajun Chen, Wei Jiang, and Yong Li. 2025. PEACE: Towards Efficient Project-Level Efficiency Optimization via Hybrid Code Editing. *arXiv preprint arXiv:2510.17142* (2025).
[49] Yingchen Tian, Yuxia Zhang, Klaas-Jan Stol, Lin Jiang, and Hui Liu. 2022. What makes a good commit message?. In *Proceedings of the 44th International Conference on Software Engineering*. 2389–2401.
[50] Xin Wang, Zhenhao Li, and Zishuo Ding. 2025. Defects4Log: Benchmarking LLMs for Logging Code Defect Detection and Reasoning. In *40th IEEE/ACM International Conference on Automated Software Engineering, ASE 2025, Seoul, Korea, Republic of, November 16-20, 2025*. IEEE, 1931–1942.
[51] Xin Wang, Zhenhao Li, and Zishuo Ding. 2026. LLM4Perf: Large Language Models Are Effective Samplers for Multi-Objective Performance Modeling. In *Proceedings of the IEEE/ACM 48th International Conference on Software Engineering (ICSE)*.
[52] Yue Wang, Hung Le, Akhilesh Deepak Gotmare, Nghi DQ Bui, Junnan Li, and Steven CH Hoi. 2023. Codet5+: Open code large language models for code understanding and generation. *arXiv preprint arXiv:2305.07922* (2023).
[53] Claes Wohlin. 2014. Guidelines for snowballing in systematic literature studies and a replication in software engineering. In *Proceedings of the 18th international conference on evaluation and assessment in software engineering*. 1–10.
[54] Yi Xiao, Van-Hoang Le, and Hongyu Zhang. 2024. free: Towards more practical log parsing with large language models. In *Proceedings of the 39th IEEE/ACM International Conference on Automated Software Engineering*. 153–165.
[55] Junjielong Xu, Ziang Cui, Yuan Zhao, Xu Zhang, Shilin He, Pinjia He, Liqun Li, Yu Kang, Qingwei Lin, Yingnong Dang, et al. 2024. Unilog: Automatic logging via llm and in-context learning. In *Proceedings of the 46th ieee/acm international conference on software engineering*. 1–12.
[56] An Yang, Anfeng Li, Baosong Yang, and et al. 2025. Qwen3 Technical Report. doi:10.48550/arXiv.2505.09388
[57] Xu Yang, Shaowei Wang, Jiayuan Zhou, and Wenhan Zhu. 2025. One-for-All Does Not Work! Enhancing Vulnerability Detection by Mixture-of-Experts (MoE). *Proceedings of the ACM on Software Engineering* FSE (2025), 446–464.
[58] Xu Yang, Wenhan Zhu, Michael Pacheco, Jiayuan Zhou, Shaowei Wang, Xing Hu, and Kui Liu. 2025. Code Change Intention, Development Artifact, and History Vulnerability: Putting Them Together for Vulnerability Fix Detection by LLM. *Proceedings of the ACM on Software Engineering* 2, FSE (2025), 489–510.
[59] Kundi Yao, Guilherme B. de Pádua, Weiyi Shang, Steve Sporea, Andrei Toma, and Sarah Sajedi. 2018. Log4perf: Suggesting logging locations for web-based systems' performance monitoring. In *Proceedings of the 2018 ACM/SPEC International Conference on Performance Engineering*. 127–138.
[60] Kundi Yao, Guilherme B. de Pádua, Weiyi Shang, Catalin Sporea, Andrei Toma, and Sarah Sajedi. 2020. Log4Perf: suggesting and updating logging locations for web-based systems' performance monitoring. *Empirical Software Engineering* 25, 1 (2020), 488–531.
[61] Xin Yin, Chao Ni, and Shaohua Wang. 2024. Multitask-based evaluation of open-source llm on software vulnerability. *IEEE Transactions on Software Engineering* (2024).
[62] Xin Yin, Chao Ni, Shaohua Wang, Zhenhao Li, Limin Zeng, and Xiaohu Yang. 2024. Thinkrepair: Self-directed automated program repair. In *Proceedings of the 33rd ACM SIGSOFT International Symposium on Software Testing and Analysis*. 1274–1286.
[63] Ding Yuan, Soyeon Park, and Yuanyuan Zhou. 2012. Characterizing logging practices in open-source software. In *2012 34th international conference on software engineering (ICSE)*. IEEE, 102–112.
[64] Chen Zhi, Jianwei Yin, Junxiao Han, and Shuiguang Deng. 2020. A preliminary study on sensitive information exposure through logging. In *2020 27th Asia-Pacific Software Engineering Conference (APSEC)*. IEEE, 470–474.
[65] Renyi Zhong, Yichen Li, Jinxi Kuang, Wenwei Gu, Yintong Huo, and Michael R. Lyu. 2025. LogUpdater: Automated Detection and Repair of Specific Defects in Logging Statements. *ACM Trans. Softw. Eng. Methodol.* (2025).
[66] Renyi Zhong, Yichen Li, Guangba Yu, Wenwei Gu, Jinxi Kuang, Yintong Huo, and Michael R Lyu. 2025. Beyond LLMs: An Exploration of Small Open-source Language Models in Logging Statement Generation. *arXiv preprint arXiv:2505.16590* (2025).
[67] Rui Zhou, Mohammad Hamdaqa, Haipeng Cai, and Abdelwahab Hamou-Lhadj. 2020. Mobilogleak: A preliminary study on data leakage caused by poor logging practices. In *2020 IEEE 27th International Conference on Software Analysis, Evolution and Reengineering (SANER)*. IEEE, 577–581.